\documentclass[fleqn,usenatbib]{mnras}

\usepackage{newtxtext,newtxmath,color}

\usepackage[T1]{fontenc}

\DeclareRobustCommand{\VAN}[3]{#2}
\let\VANthebibliography\thebibliography
\def\thebibliography{\DeclareRobustCommand{\VAN}[3]{##3}\VANthebibliography}

\usepackage{graphicx}	
\usepackage{amsmath}	
\usepackage{anyfontsize}
\usepackage{float}

\newcommand{\z}[1]{\mbox {ZZ PsA}}

\def\p0{\phantom{0}}

\usepackage{xcolor}

\title[Contact binary evolution and orbital stability]{Low Mass Contact Binaries: Orbital Stability at Extreme Low Mass Ratios.}

\author[S. S. Wadhwa et al.]{Surjit S. Wadhwa,$^{1}$\thanks{E-mail: 19899347@student.westernsydney.edu.au}
Natália R. Landin,$^{2}$
Bojan Arbutina,$^{3}$
Nicholas F.\,H. Tothill,$^{1}$
\newauthor Ain Y. De Horta,$^{1}$
Miroslav D. Filipovi\'c,$^{1}$ 
Jelena Petrovi\'c$^{4}$ and
Gojko Djura\v sevi\'c$^{4}$
\\
$^{1}$School of Science, Western Sydney University, Locked Bag 1797, Penrith, NSW 2751, Australia\\
$^{2}$Universidade Federal de Viçosa, Campus UFV Florestal, CEP 35690-000 Florestal, MG, Brazil\\
$^{3}$Department of Astronomy, Faculty of Mathematics, University of Belgrade, Studentski trg 16, 11000 Belgrade, Serbia\\
$^{4}$Astronomical Observatory, Volgina 7, 11060 Belgrade, Serbia\\
}

\date{Accepted XXX. Received YYY; in original form ZZZ}

\pubyear{2023}

\begin{document}
\label{firstpage}
\pagerange{\pageref{firstpage}--\pageref{lastpage}}
\maketitle

\begin{abstract}
With the ever-increasing number of light curve solutions of contact binary systems increasing number of potential bright red nova progenitors are being reported. There remains, however, only one confirmed event. In the present study we undertake a comprehensive review of orbital stability of contact binary systems considering the effects of the stellar internal composition (metallicity) and age on the evolution of the gyration radius and its effect on the instability mass ratio of contact binaries. We find that both metallicity and age have an independent effect on orbital stability with metal poor and older systems being more stable. The combined effects of age and metallicity are quite profound such that for most systems with primaries of solar mass or greater which are halfway or more through the main sequence lifespans have instability mass ratio at levels where the secondary component would be below the hydrogen fusion mass limit. We find that from the currently available solutions we cannot confidently assign any system as unstable. Although we identify 8 potential red nova progenitors all have methodological or astrophysical concerns which lower our confidence in designating any of them as potential merger candidates. 

\end{abstract}




\section{Introduction}
 \label{sec:intro}

W Ursa Majoris binary systems usually comprise of a low mass (spectral class F to K) primary with an even lower mass secondary. Due to their proximity the components are distorted, and both fill their Roche lobes leading to mass exchange between the components and the development of a common atmosphere. The systems are also referred to as contact binaries or over-contact systems. Contact binaries are quite common with estimates as high as 1 in 500 stars in the Galactic disk \citep{2006MNRAS.368.1319R}. The primary behaves like a main sequence star of the corresponding mass while the secondary has a larger radius and is more luminous than a matched main sequence counterpart due to expansion beyond its Roche lobe and the energy transfer from the primary through the common atmosphere. The presence of the common atmosphere results in almost identical photospheric temperatures for both components near the main sequence temperature of the primary component \citep{2013MNRAS.430.2029Y}. The median orbital period is near 0.35 days although it can range from about 0.2 to over 1 day \citep{2021ApJS..254...10L}. 

Confirming that the transients termed red novae were the likely result of contact binary merger events \citep{2011A&A...528A.114T} has led to greater interest in the study of orbital stability of such systems. Although there is only a single confirmed merger event, that of V1309 Sco \citep{2011A&A...528A.114T}, other galactic transients such as V4432 Sgr \citep{1999AJ....118.1034M}, V838 Mon \citep{2002IAUC.7785....1B} and OGLE2002-BLG-360 \citep{2013A&A...555A..16T} also likely represent merger events; however, the pre-transient progenitors are as yet not confirmed contact binaries.  Estimates of the galactic frequency of such events are as high as once every two to three years \citep{2014MNRAS.443.1319K}, although observable events may be less frequent. Theoretical developments over the past couple of decades have demonstrated that orbital instability leading to merger is critically dependent on the observed mass ratio between the components \citep{2007MNRAS.377.1635A, 2009MNRAS.394..501A, 2010MNRAS.405.2485J, 2021MNRAS.501..229W, 2024SerAJ.208....1A}. The instability mass ratio itself is critically linked to the internal structure of the primary component and hence influenced by metallicity \citep{2024MNRAS.527....1W} and potentially by the evolution of the system \citep{1995ApJ...438..887R, 2006MNRAS.369.2001L, 2010MNRAS.405.2485J}.

The present work presents a comprehensive overview of the orbital stability of contact binaries taking full account of metallicity and main sequence evolution. The instability mass ratio is influenced by two variables: one physical and one structural. The mass ratio is essentially proportional to the structural parameter called the gyration radius ($k=\sqrt{I/MR^2})$, where $I$ is the moment of inertia, $M$ the mass and $R$ the radius) of a star and weakly influenced by the degree of contact or fillout factor($f$). We model the gyration radii of low mass stars from 0.6\,$\rm M_{\odot}$ to 1.4\,$\rm M_{\odot}$ taking into account rotation and binary tidal effects at different metallicities from zero-age main sequence (ZAMS) to terminal-age main sequence (TAMS). We apply them to the mass ratio instability equations from \citep{2024SerAJ.208....1A} to derive the instability mass ratios for contact binaries with primaries ranging from 0.6\,$\rm M_{\odot}$ to 1.4\,$\rm M_{\odot}$ for various metallicities ranging from $-1.25\leq \rm [Fe/H] \leq 0.5$ and over a timescale from ZAMS to TAMS. The instability mass ratios were determined at marginal contact ($f=0$) and full over contact ($f=1$). The process yields an instability mass ratio range at various metallicities for the main sequence timescale for low mass contact binaries.

Finally, we reviewed the cataloged contact binary light curve solutions from \citet{2021ApJS..254...10L} and carried out literature search up to the end of March 2024 to compile a list of potential red nova progenitors meeting the revised criteria accounting for metallicity and age. 

\section{Orbital Stability}
We base our analysis of orbital stability on the recent updated criteria by \citep{2024SerAJ.208....1A}. Full derivation of the instability mass ratio equations at marginal contact ($f = 0$) and full over-contact ($f = 1$) can be found in the cited reference. As suggested by \citet{2009MNRAS.394..501A} and modeled by \citep{2024MNRAS.527....1W}, at the onset of instability the mass of the secondary is well below the mass where a star becomes fully convective, and it is reasonable to adopt the $n = 1.5$ polytrope value ($k_2 = 0.4527$) as the gyration radius value for the secondary component. Although initial investigators employed fixed values for the gyration radius of the primary, it is now clear that individually modeled values for the gyration radius of the primary have significant influence on the instability mass ratio \citep{2021MNRAS.501..229W}.

The updated instability mass ratio criteria from \citet{2024SerAJ.208....1A} when combined with the fixed value of $k_2$ can be approximately reduced to two simple quadratic relations linking the gyration radius of the primary ($k_1$) to the instability mass ratio at marginal ($f=0$) and full over-contact ($f=1$) as follows:

\begin{equation}
    q{\rm_{inst}} = 0.24501\times k_{1}^2 + 1.15793\times k_1 + 0.00696 \\    (f=1)
\end{equation} and
\begin{equation}
    q{\rm_{inst}} = -0.64933\times k_{1}^2 + 1.11635\times k_1 + 0.00704 \\    (f=0).
\end{equation}

The instability mass ratio moves in line with the gyration radius of the primary. The smaller the gyration radius the smaller the instability mass ratio. Internal structural constants have been modeled for some time and metallicity and age have a significant influence on the gyration radius given the evolving density and radius profiles \citep{2007A&A...467.1389C}. In the next section we will systematically explore the variation in the gyration radius of a solar mass star incorporating rotation, binary tidal distortions, metallicity and evolving age. We then look at effects of metallicity and age separately on the instability mass ratio and finally we examine the instability mass ratio with combined effects of metallicity and age. We extend our modeling to cover systems with primary components ranging from $0.6\rm M_{\odot}$ to $1.4\rm M_{\odot}$, for metallicities $\rm -1.25\leq\rm[Fe/H]\leq0.5$ and ZAMS to TAMS ages.

\subsection{Model}

The gyration radii of low-mass (from 0.6 to 1.4\,M$_{\odot}$) stars were modeled using the
\textsc{aton} stellar evolution code \citep{landin06,2009A&A...494..209L} and taking into account the
effects of rotation and tidal distortions. We generated 8 sets of models with different metallicities in the range -1.25$\leq$${\rm[Fe/H]}$$\leq$0.5.
The solar metallicity (${\rm[Fe/H]}$=0) corresponds to the adopted solar chemical
composition (X = 0.7125 and Z = 0.0175) from \citet{anders89}. We assumed that the elements are mixed
instantaneously in convective regions. The convective transport of energy is treated
according to the conventional Mixing Length Theory \citep{vitense58}, with the parameter that
represents the convection efficiency set as $\alpha$=1.5. Surface boundary conditions were
obtained from grey atmosphere models and the matching between the surface and the interior
is made at the optical depth $\tau$=2/3. We used the equations of state from \citet{rogers2}
and \citet{mihalas} and the opacities reported by \citet{rogers1} and \citet{1994ApJ...437..879A}.
Our models were generated by considering rigid body rotation
without taking into account internal angular momentum redistribution and angular momentum
loss by stellar magnetized winds \citep{mendes99}.
The initial angular momentum of each model was obtained according to the \citet{kawaler87} relation
\begin{equation}
J_{\rm kaw}=(1.562 \pm 0.504)\times 10^{50} \left( \frac{M}{M_{\odot}} \right) ^{(0.985 \pm 0.140)}
~~~{\rm{g~cm^2~s^{-1}.}}
\label{eqkaw}
\end{equation}
The effects of tidal distortions due to a companion star (in a binary system) were computed as detailed in \citep{2009A&A...494..209L}. 
%

\subsection{Gyration radius, instability mass ratio, metallicity and age}

Figure 1 illustrates the gyration radius of a solar mass star from ZAMS to TAMS at various metallicities. The main points to note are:
\begin{itemize}
    \item Regardless of age, the gyration radius is smaller at lower metallicity;
    \item The gyration radius declines throughout the main sequence but is stable in the first gigayear; 
    \item The reduction in gyration radius with age is more pronounced in stars with lower metallicity.
\end{itemize}
We must keep in mind that metallicity has an effect on main sequence lifetime (ZAMS to TAMS) such that lower metallicity systems have a shorter lifespans and higher metallicities spend much longer on the main sequence \citep{1996A&AS..115..339C, 1999A&AS..135..405C}. In the case of a solar mass star the time to TAMS can range from just over 5 Gyr at $\rm[Fe/H] = -1.25$ to over 20 Gyr at $\rm[Fe/H] = 0.5$. The trend is similar for primaries from 0.6$\rm M_{\odot}$ to 1.4$\rm M_{\odot}$ but less pronounced in stars of lower mass. 

\begin{figure*}
    \centering
    \includegraphics [width=\textwidth]{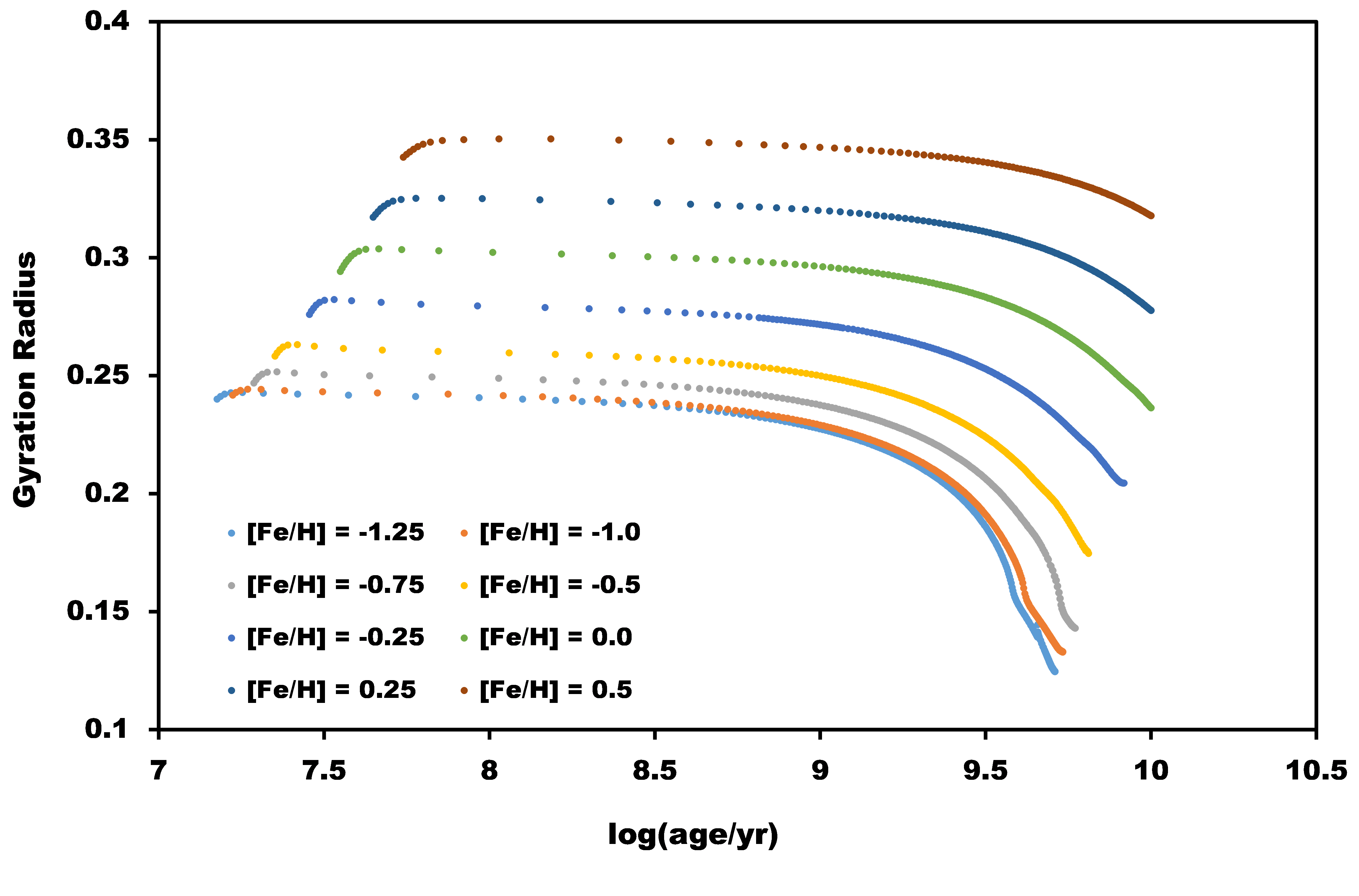}
    \caption{Gyration radius for a solar mass star from ZAMS to TAMS for various metallicities}
    \label{fig:kagemetal}
\end{figure*}


We next look at the effects of evolutionary age and metallicity on the instability mass ratio ($q_{inst}$) range. Figure 2 illustrates $q_{inst}$ at $f=0,1$ for a solar mass star ($\rm[Fe/H] = 0.0$) from 1 Gyr to 10 Gyr. At 1 Gyr $q_{inst}$ range is $0.107 - 0.126$; however, this reduces to $0.073 - 0.082$ at 10 Gyr. It is interesting to note that at solar metallicity a contact binary system with a solar mass primary would become unstable near TAMS with the mass of the secondary potentially below that required for sustained hydrogen fusion. Modeling the same system with metallicity $\rm[Fe/H] = -1.25$ the instability mass ratio at 1 Gyr is 0.068 - 0.076 and at the TAMS ($\approx$5 Gyr) it falls to 0.024 - 0.026. So regardless of age a very low metallicity solar mass primary system will only reach instability when the secondary is unable to maintain hydrogen fusion and at near TAMS the mass of the secondary at the onset of instability more closely resembles a large gaseous planet. At a higher metallicity ($\rm[Fe/H] = 0.5$), note the main sequence lifespan is almost 4 times longer than at metallicity $\rm[Fe/H] = -1.25$, the instability mass ratio at 1 Gyr is 0.141 - 0.173 and at the TAMS (older than current age of the Universe) this reduces to 0.100 - 0.117. Clearly there is profound effect of metallicity and evolutionary aging on the orbital stability of contact binaries. If one considers a low mass ratio as a marker of stability, then in general low metal content systems are more stable. A review of The Large Sky Area Multi-Object Fiber Spectroscopic Telescope (LAMOST) catalog of contact binaries \citep{2020RAA....20..163Q} shows that over 70\% of the objects have metallicities below solar metallicity.
In addition, mass loss in contact binaries through stellar winds, and certainly mass transfer from secondary to primary, in addition to the angular momentum loss, does complicate stability analysis, but since it will probably widen the orbit, it would add up to the increasing stability offered by changes in the internal structure (lower gyration radius) as contact binary systems age, and they in fact can become more stable, reducing the potential of being merger candidates.

\begin{figure*}
    \centering
    \includegraphics [width=\textwidth]{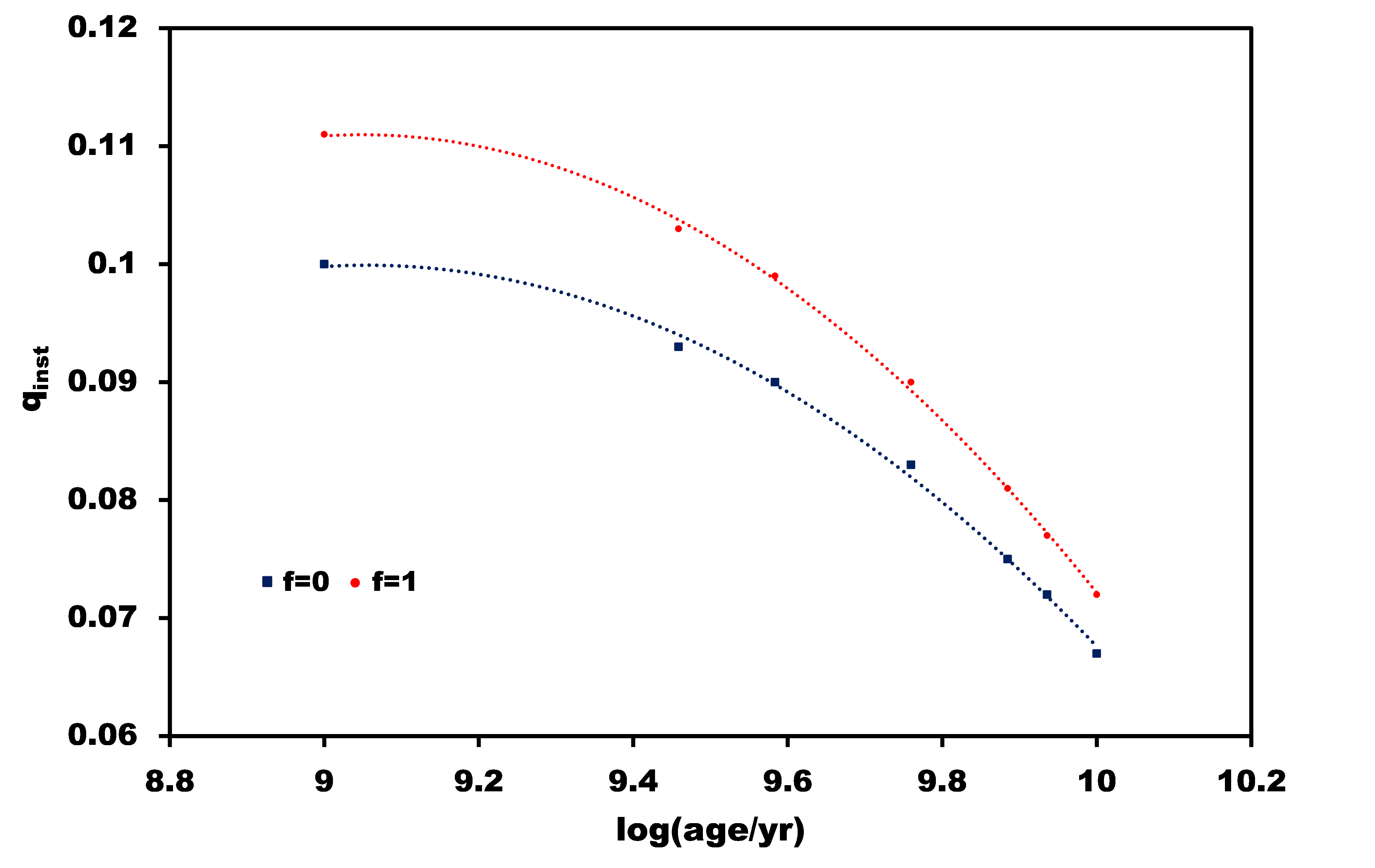}
    \caption{Reduction in the instability mass ratio from 1Gyr to 10Gyr for a contact binary with a solar mass primary at solar metallicity.}
    \label{fig:qiage}
\end{figure*}

\section{Instability Criteria - An Update}
It is widely accepted that the primary component of a low mass contact binary system behaves like a main sequence star of similar mass. It is, however, impossible to know how old the primary component is as pre-contact evolution of the components is unknown. If one accepts the proposed origin of contact binaries from semi-detached systems where the designation of the primary and secondary have switched due to mass transfer during the semi-detached stage \citep{2006AcA....56..199S}, then it is possible that the system as a whole could be quite old; however, the "higher mass" phase of the current primary may not be that old. To this end we propose an update for estimating the orbital stability of contact binary systems.

Firstly, the mass of the primary component needs to be determined as accurately as possible. The method of choice would be through radial velocity measurements; however, this is rarely possible. We suggest that methods that do not involve the use of the assigned temperature of the primary component be used. We say this because various calibrations for the temperature can yield very wide estimates \citep{2023SerAJ.207...21W} and even small variations in the assigned temperature can result in a significant difference in the estimate of the mass. Using simple black body calculations, a change in temperature of $250\rm K$ can result in a greater than 5\% change in the mass estimate. We prefer direct observational methods, with accurately determined distances and extinctions or colour calibrations where the magnitude estimates in different bands have been acquired simultaneously \citep{2023PASP..135g4202W}. Secondly, the metallicity should be determined spectroscopically or potentially through the GAIA data release 3 \citep{2022A&A...658A..91A}. Note that the photometric GAIA metallicity needs to be corrected as described in \citep{2023PASP..135g4202W}. Lastly, we propose that the gyration radius of the primary should be estimated at a set age of half the TAMS age or 10 Gyr where half TAMS age exceeds 10 Gyr. This is a somewhat arbitrary set point; however, as stated above it is impossible to know the age of the primary in its current form and as such the selection, we feel, is a reasonable compromise.

Below we present a set of associations for determining the instability mass ratio range for contact binary systems with primaries ranging from $0.6\rm M_{\odot}$ to $1.4\rm M_{\odot}$ (in steps of 0.1\,$\rm M_\odot$) and metallicities from $\rm -1.25\leq\rm[Fe/H]\leq0.5$ in steps of $\rm[Fe/H] = 0.25$.

\subsection{Instability Mass ratio, Metallicity and Age}
Figure 3 illustrates the change in the instability mass ratio range ($f=0,1$) at $\rm[Fe/H] = 0.0$ for contact binaries with primaries from $0.6\rm M_{\odot}$ to $1.4\rm M_{\odot}$. The age of the system was set at half the TAMS age or 10 Gyr if half TAMS age was greater than 10 Gyr. As can be seen from figure 3 there is a good quadratic relation between the mass of the primary and the instability mass ratio, as previously shown by \citep{2021MNRAS.501..229W}

\begin{figure*}
    \centering
    \includegraphics [width=\textwidth]{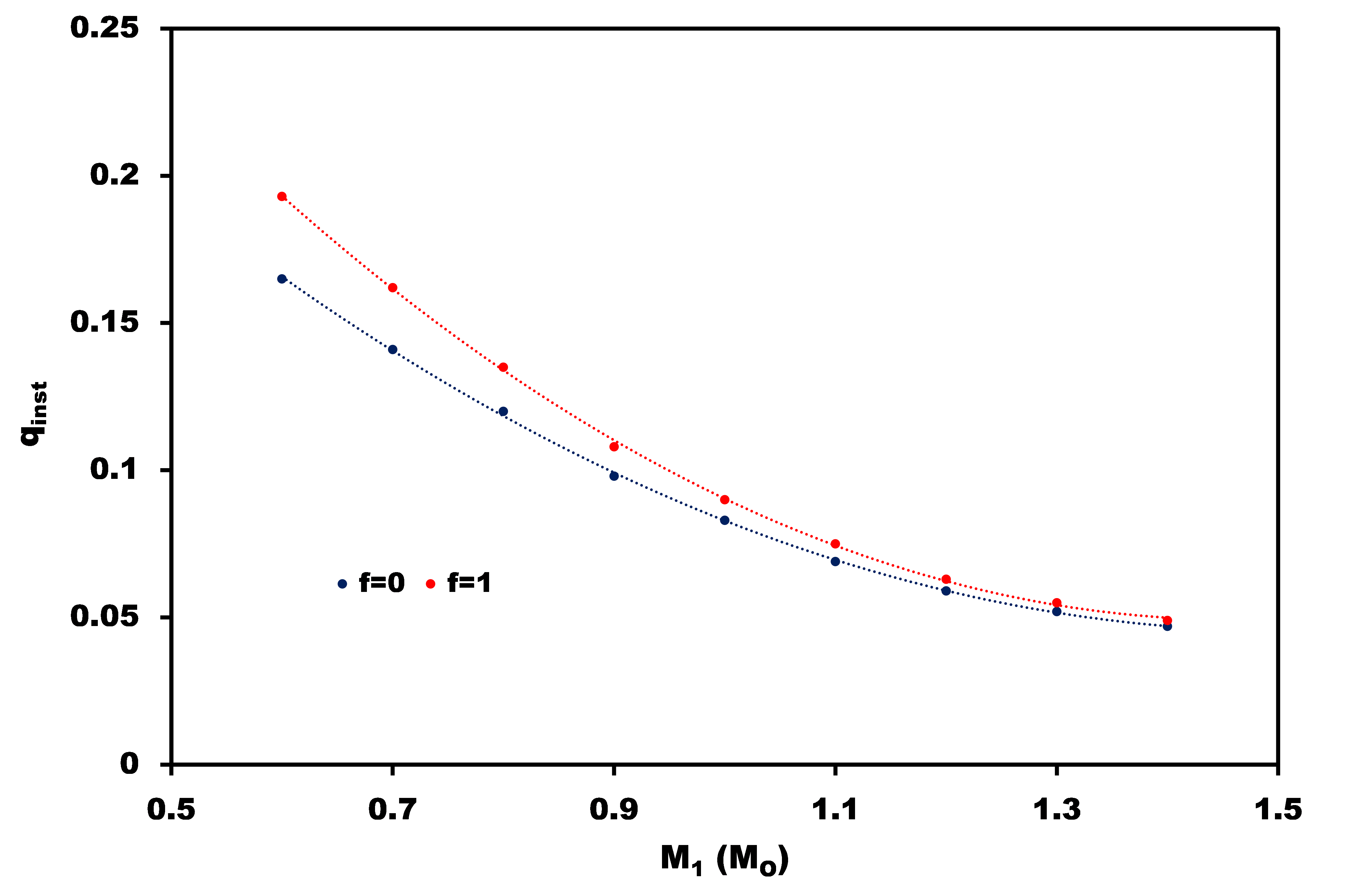}
    \caption{Instability mass ratio, for f=0 and f=1, at solar metallicity and half TAMS age or 10 Gyr if half TAMS age was greater than 10 Gyr for contact binary systems with primaries from $0.6\rm M_{\odot}$ to $1.4\rm M_{\odot}$. There is strong quadratic relationship between the mass of the primary and the instability mass ratio range. Quadratic relationships continue for the full range of metallicities.}
    \label{fig:qinst}
\end{figure*}
\begin{equation}
    \rm q_{inst} = a\rm M_1^2 + b\rm M_1 + c.
\end{equation}

The relationship holds true for the full metallicity range and we provide the quadratic coefficients ($a, b, c$) for metallicities $\rm -1.25\leq\rm[Fe/H]\leq0.5$ in steps of $0.25$ in Table 1. To determine the progenitor status of a system, we propose to check the instability mass ratio range at the closest metallicity value. For the rare systems that are say within 0.05 of the halfway of the modeled metallicity, one can adopt the mean value of the instability mass ratio range for metallicities on either side. 
\begin{table}
\caption{Quadratic coefficients (a, b, and c) at f=1 and f=0 for the full range of metallicities with the age set at half TAMS or 10Gyr whichever is the lesser. The correlation coefficients ($R^2$) in all cases range from 0.97 to 0.99.}
\begin{tabular}{ccclclclc}
\textbf{[Fe/H]}  & \textbf{f=1}                     & \textbf{f=0}                     \\
-1.25 & 0.3324, -0.7400, 0.4737 & 0.2732, -0.6277, 0.4077 \\
-1.00 & 0.2573, -0.6080, 0.4087  & 0.2306, -0.5451, 0.3685 \\
-0.75 & 0.2828, -0.6914, 0.4678 & 0.2373, -0.5822, 0.4003 \\
-0.50 & 0.2640, -0.6681, 0.4685  & 0.2162, -0.5528, 0.3968 \\
-0.25 & 0.2218, -0.6057, 0.4591 & 0.1750, -0.4872, 0.3818 \\
0.00  & 0.1948, -0.5686, 0.4642 & 0.1648, -0.4418, 0.3780 \\
0.25  & 0.1753, -0.5426, 0.4743 & 0.1222, -0.4005, 0.3757 \\
0.50  & 0.1313, -0.4624, 0.4605 & 0.0794, -0.3199, 0.3560
\end{tabular}
\end{table}


\section{Red Nova Progenitors - A Survey}
In this section we review the catalog of contact binary solutions (648 systems) from \citet{2021ApJS..254...10L} which covers up to the end of 2020. In addition, we performed a literature search for published contact binary solutions from 2021 to the end March 2024. We accepted the authors estimate of the mass of the primary, regardless of the methodology. As our modeling is for systems from $0.6\rm M_{\odot}$ to $1.4\rm M_{\odot}$ we excluded all solutions where the estimated mass of the primary was outside this range. Similarly, if the metallicity was not available or not within our modeled range the system was excluded. As noted above, the instability mass ratio is higher with low mass high metallicity primaries and the maximum instability mass ratio for our select systems is 0.226 ($M_1 = 0.6\rm M_{\odot}$ and $f=1$) and as such all solutions with an estimated mass ratio above 0.226 were excluded. Based on the estimated mass of the primary, we next determined the systems potential maximum instability mass ratio, i.e at $f=1$ and $\rm [Fe/H] = 0.5$. If the mass ratio was in excess of the maximum potential instability mass ratio the system was excluded. We then estimated the metallicity of the system based on either the corrected GAIA photometry estimate \citep{2023PASP..135g4202W} or spectroscopic estimate from the seventh data release (DR7) of the LAMOST survey \citep{2020RAA....20..163Q}. Where more than one estimate was available from the LAMOST survey we selected the estimate based on the higher signal to noise ratio. Next, we determined the instability mass ratio range at the estimated metallicity as described above. 

\subsection{Specific Systems}
Table 2 lists the 8 systems that meet our criteria for potential instability. We discuss each of the systems below with comments regarding the estimate of the mass of the primary and comparison of the estimated mass from GAIA \citep{2022A&A...658A..91A} and TESS input catalogs \citep{2021arXiv210804778P} or other possible methodology issues which reduce our confidence for potential instability. 

\begin{table}
\caption{Initial list of 8 potential red nova progenitors. M\textsubscript{1} and q are the published mass of the primary and mass ratio. The [Fe/H] column lists the metallicity for each system as determined through LAMOST (L) or GAIA (G). The q\textsubscript{inst} range is the instability mass ratio range for the published value of M\textsubscript{1} and determined metallicity with age set at half TAMS.}
\begin{tabular}{lclclclclc}
\textbf{Name}                  & \textbf{M\textsubscript{1}}    & \ \textbf{  \ q}      & \textbf{[Fe/H]}    & \textbf{q\textsubscript{inst} range}   \\
V1187 Her& 1.15&0.044&-0.20 L&0.053 - 0.056\\
NSVS 780649           & 0.79  & 0.098  & -0.40 G & 0.095 - 0.105 \\
GSC 02800-1387        & 0.85  & 0.115  & 0.09 L  & 0.109 - 0.122 \\
TYC 7281-269-1        & 1.01  & 0.097  & 0.31 G  & 0.096 - 0.105 \\
ASAS J170715-5118.7   & 1.22  & 0.072  & 0.50 G   & 0.084 - 0.092 \\
WISE J185503.7+611804 & 1.00  & 0.051  & 0.00 G  & 0.083 - 0.090 \\
WISE J141530.7+592234 & 1.04  & 0.055  & 0.21 G  & 0.071 - 0.077 \\
TYC 4002-2628-1       & 1.11  & 0.048  & 0.34 G  & 0.091 - 0.099\\
\end{tabular}
\end{table}


\subsubsection{V1187 Her}
V1187 was observed by \citet{2019PASP..131e4203C} with the reported mass ratio of only 0.044. This estimate was based on photometric light curve solutions where the contribution of the third light was modeled at near zero. Based on the spectral classification of F8.5 the estimated mass of the primary is $1.15\rm M_{\odot}$. There is, however, recent evidence based on analysis of TESS photometry and echelle spectra to suggest that the system is contaminated by a significant third light contribution and recent estimates of the mass ratio of the system are as high as 0.16 \citep{2022AAS...24020505C, 2024AAS...24342302C}. Given the uncertainty in the estimated mass ratio for the system we consider the system to be likely stable.

\subsubsection{NSVS 780649}
NSVS 780649 was observed by \citet{2021RAA....21..225P} with reported mass ratio of 0.098 and mass of the primary as $0.79\rm M_{\odot}$. The light curve solution is based on a spotted profile which is prone to the problem of non-uniqueness \citep{1999NewA....4..365E}  so it is possible the geometric solution may differ from the one published. In addition, the estimated mass of the primary is significantly less than the values obtained with GAIA ($1.00\rm M_{\odot}$) and TESS ($0.98\rm M_{\odot}$). We cannot definitively exclude the system; however, we are cautious with respect to its instability.




\subsubsection{GSC 02800-1387}
GSC 02800-1387 was observed by \citet{2022NewA...9701862P} with reported mass ratio of 0.115 and mass of the primary $0.85\rm M_{\odot}$. The system has a similar profile to NSVS 780649 in that a spotted solution is used to estimate the geometric parameters and the estimated mass of the primary is considerably less than the GAIA ($1.01\rm M_{\odot}$) and TESS ($1.02\rm M_{\odot}$) estimates. Again, we are cautious as to the instability classification for this system.

\subsubsection{TYC 7281-269-1}
TYC 7281-269-1 was observed by \citet{2022RAA....22j5009W} with a reported mass ratio of 0.097 and mass of the primary as $1.01\rm M_{\odot}$. The GAIA ($1.25\rm M_{\odot}$) and TESS ($1.22\rm M_{\odot}$) estimates for the mass are significantly higher and we remain cautious regarding the instability status of this system.

\subsubsection{ASAS J170715-5118.7}
ASAS J170715-5118.7 was observed by \citet{2023PASP..135g4202W} with a reported mass ratio of 0.072 and mass of the primary as $1.22\rm M_{\odot}$. The GAIA ($1.32\rm M_{\odot}$) and TESS ($1.28\rm M_{\odot}$) estimates for the mass are somewhat higher then reported which suggests a degree of caution. Another interesting feature is the calculated mass of the secondary from the reported mass ratio. Assuming the primary has a mass of $1.22\rm M_{\odot}$ then from the mass ratio the mass of the secondary would be $0.088\rm M_{\odot}$. The value is very close to the lower limit at which hydrogen fusion is possible and this is reflected in the relatively poor thermal contact with the secondary almost 500K cooler than the primary.

\subsubsection{TYC 4002-2628-1, WISE J141530.7+592234 and WISE J185503.7+611804}

All three systems discussed in this section share extreme mass ratios such that, based on the estimated masses of the primary components, they all have secondaries with masses well below the hydrogen burning mass. If these were non-evolved isolated objects, they would be brown dwarfs. However, these are expanded stars (filling their respective Roche lobes) in thermal contact with the primary, and the mass threshold for the secondary to continue hydrogen fusion may not be the same as the mass threshold required for a newborn star to ignite hydrogen fusion for the first time. Although the internal structure of the secondary is uncertain, it must somehow account for the energy transfer. Examples of main sequence stars that have lost mass through expansion beyond the Roche lobe such that their mass is below the hydrogen burning limit have been identified in the case of cataclysmic binaries and show little intrinsic luminosity \citep{2003MNRAS.340..264L, 2022MNRAS.509.5086W}.

The widely accepted Lucy model of contact binaries \citep{1968ApJ...151.1123L} is predicated on significant transfer of energy/luminosity from the primary to the secondary. To maintain the temperature of the common atmosphere the extent of energy transfer will increase with decreasing contribution from the secondary. Although it is not possible to directly measure the energy transfer, we can safely say that the total observed luminosity ($\rm L_o$) of the system cannot be more than the sum of the nuclear luminosities ($\rm L_n$) of the components. Both \citet{2001OAP....14...33K} and \citet{2009MNRAS.396.2176J} have confirmed this for a large sample of contact binaries. We examine the three systems with respect to their observed and theoretical luminosity relations.

TYC 4002-2682-1 was observed by \citet{2022MNRAS.517.1928G} with reported mass ratio of 0.0482 and masses of the components as $\rm M_1 = 1.11\rm M_{\odot}$ and $\rm M_2 = 0.054\rm M_{\odot}$. They reported the luminosities of the components as $\rm L_1 = 2.49\rm L_{\odot}$ and $\rm L_2 = 0.16\rm L_{\odot}$. \citet{2009MNRAS.396.2176J} showed that a check of congruence between observed and nuclear luminosities could be performed by examining the log of the total observed and nuclear luminosities as follows: [$\rm log( L_{o1} + L_{o2}) = log (L_{n1} + L_{n2})$]. For consistency we use the same methodology as \citet{2009MNRAS.396.2176J} to determine the nuclear luminosity of the primary component while we use the nuclear luminosities of brown dwarfs from \citet{2015ApJ...810..158F} to estimate the secondary contribution. The observed luminosity [$\rm log( L_{o1} + L_{o2})$] is estimated as 0.42, while the nuclear luminosity [$\rm log( L_{n1} + L_{n2})$] is significantly lower (at 0.17). As the observed luminosity is significantly greater than the nuclear one this suggests possible additional contribution.

In addition, the reported absolute parameters do not correlate with the observational data. Review of the V band light curve and the magnitude of the comparison star yield an apparent brightest (both components) magnitude of 11.49. The distance corrected extinction for the system is high at 0.32 magnitude. The GAIA distance for the system is estimated at 452pc yielding a combined total luminosity of $\rm 5.47L_{\odot}$. Assuming the light curve solution for the luminosity ratio yields the luminosity of the primary as $\rm 5.11L_{\odot}$, corresponding to a mass of the primary as $\approx \rm 1.45M_{\odot}$. This value is close to the GAIA estimated mass of $\rm 1.34M_{\odot}$. Adopting the revised luminosities the system still does not conform to the observed-nuclear congruence.

Finally, a review of the GAIA spectroscopic data suggests that the primary component may be a spectroscopic binary with a period of 1055 days. Given the discrepancy in the observed and nuclear luminosities, non-correlative observed and reported parameters and the potential binary status of the primary we conclude that there is likely a significant third light contribution and the reported mass is ratio unreliable.

WISE J141530.7+592234 was observed by \citet{2023PASP..135d4201G} with a reported mass ratio of 0.055, mass of the primary as  $1.04\rm M_{\odot}$, mass of the secondary as $0.057\rm M_{\odot}$, and luminosities of the components as $\rm L_1 = 1.98L_{\odot}$ and $\rm L_2 = 0.2L_{\odot}$. Using the same methodology as above we have $\rm log( L_{o1} + L_{o2}) = 0.34$ and $\rm log( L_{n1} + L_{n2}) = 0.07$, again indicating the observed luminosity in excess of the nuclear luminosity.

As with TYC 4002-2682-1 there is a discrepancy between the observational data and the reported absolute parameters. The extinction corrected apparent brightest magnitude for the system is 13.18, yielding an observed total luminosity of $\rm 1.87L_{\odot}$. As noted by \citet{1968ApJ...153..877L}, unlike main sequence stars, contact binaries follow a simpler mass luminosity relation, $\rm L_1/L_2 = q^{0.92}$. Adopting the revised total luminosity and main sequence luminosity for the primary star ($\rm 1.2L_{\odot}$) yields an estimated mass ratio significantly greater than the reported mass ratio. Again, given the obvious discrepancies, we suspect there is likely third light contribution to the observed light curve.

The situation is the same with WISE J185503.7+611804\footnote{The original manuscript records an incorrect declination of +592234.} observed by \citep{2023MNRAS.521...51G} with reported mass ratio of 0.0514, mass of the primary as $\rm 0.995M_{\odot}$, mass of the secondary as $\rm 0.0517M_{\odot}$ and component luminosities as $\rm L_1 = 1.32L_{\odot}$ and $\rm L_2 = 0.1L_{\odot}$. The log values for the combined reported and nuclear luminosities are 0.15 and 0.005 respectively, again showing significantly higher observed luminosity relative to the nuclear luminosity. The difference between the reported and the observed total luminosity in this case is not as great as in the above two examples with the light curve estimated total luminosity as $\rm 1.26L_{\odot}$. Adopting the main sequence luminosity for a solar mass star, the estimated mass ratio based on Lucy's relationship is significantly higher than the reported value. Significant third light contribution is again suspected.

\subsubsection{Borderline cases}
For completeness, we list in Table 3 a number of systems where the reported mass ratio is near the instability mass ratio. These, however, are not discussed further. We emphasize again that our modeling is limited to $M_1 < 1.4M_\odot$, so a number of systems with higher estimated masses of the primaries were excluded from the analysis. For example, all systems reported in \citet{2016PASA...33...43S} have estimated primary masses larger than 1.4 $M_\odot$, and the same is true for the half of the systems reported by \citep{2022MNRAS.512.1244C}. For the other half (15 systems)  our analysis shows that all are stable and as such were not reported. Some extreme low-mass ratio systems with higher mass primary, such as IP Ly ($q=0.055$, \citep{2023RAA....23h5013Y}), KIC 4244929 and KIC 9151972 ($q=0.059$, \citep{2016PASA...33...43S}) seem to be stable even for ZAMS gyration radius \citep{2024SerAJ.208....1A}.


\section{Discussion and Conclusions}

The recognition of V1309 Sco transient as a red nova and likely merger event of contact binary components combined with identification of literally hundreds of thousands of contact binaries from various surveys has ignited the field of contact binary light curve modeling and identification of red nova progenitors. Regardless of the number of new contact binary identified, V1309 Sco remains the only confirmed merger case. Recent theoretical and practical refinements \citep{2021MNRAS.501..229W, 2022JApA...43...94W} were envisaged to assist in the identification of potential red nova progenitors and many investigators (see citations in section 4.1) used the updated criteria with confident predictions of unstable systems.

\begin{table}
\caption{List of borderline systems with mass ratios near the instability mass ratio range. 1: \citep{2021ApJ...922..122L}, 2: \citep{2022NewA...9701862P}, 3: \citep{2022RAA....22j5009W}, 4: \citep{2020PASJ...72..103L}, 5: \citep{2015PASJ...67...74L}, 6: \citep{2023PASP..135g4202W}
}
\begin{tabular}{lclclc}
\textbf{Name}                  & \textbf{Mass Ratio (q)}    & \ \textbf{q\textsubscript{inst} range}      \\
VSX	J082700.8+462850$^1$ & 0.055&0.051 - 0.054\\
V755Cep$^2$           & 0.136  & 0.117 - 0.131  \\
ASAS J082151-0612.6$^3$        & 0.097  & 0.081 - 0.089 \\
KIC	8432859$^4$ & 0.090  & 0.078 - 0.084 \\
V1222Tau$^5$   & 0.112  & 0.091 - 0.100 \\
ASAS J100101-7958.6$^6$ & 0.138  & 0.111 - 0.124 \\

\end{tabular}
\end{table}

The most critical parameters determining orbital stability are the mass ratio and the gyration radius of the primary. The fill-out (degree of contact) plays a minor but crucial role. Where the light curve displays a total eclipse the mass ratio and fill-out can be reliably obtained from the light curve solution; however, care must be taken to avoid local degenerated solutions and potential third light contributions should be adequately addressed. As the gyration radius is dependent on the internal structure of the star, it was adjusted in the earlier stages to match the observations rather than formally modeled \citep{1995ApJ...444L..41R}. When the gyration radius was modeled \citep{1988AJ.....95.1895R, 2009A&A...494..209L, 2010MNRAS.405.2485J} the results were to some extent overlooked by researchers even though they pointed to a significantly different understanding to the prevailing view of a universal minimum mass ratio for all contact binaries. \citet{2021MNRAS.501..229W} refined the orbital stability criteria and using previously modeled gyration radii values for rotating and tidally distorted stars of different mass (at the ZAMS) and showed that there was no one universal instability mass ratio, in fact, each system has its own instability mass ratio. 

As noted above the gyration radius is dependent on the internal structure of the star and all stars have different composition (metallicity) so it was clear that the role of metallicity on the instability mass ratio needed exploration. Some very preliminary work \citep{2010Ap&SS.329..283J} indicated that the effect on the "minimum" mass ratio from changing metallicity is likely to be not insignificant. \citet{2024MNRAS.527....1W} comprehensively modeled the gyration radii of low mass contact binary systems for various metallicities from $\rm -1.25\leq\rm[Fe/H]\leq0.5$ and showed that regardless of the mass of the primary component the instability mass ratio was lower for lower metallicities, i.e. metal poor systems were more likely to be stable. A review of the LAMOST spectra \citep{2020RAA....20..163Q} suggests that over 70\% of over 7000 surveyed contact binaries have metallicities below solar metallicity.

Internal structure of the star changes considerably even during main sequence evolution and as such it is likely the gyration radius is likely to change and in turn have an influence on orbital instability. Just as in the case of metallicity some earlier works indicated a potential profound effect especially in the later parts of the main sequence lifespan \citep{2010MNRAS.405.2485J}. In the present study, for the first time as far as we are aware, we model the gyration radii of rotating, tidally distorted stars at various metallicities and for the entire main sequence (ZAMS to TAMS) lifespan. We can confirm the previous finding that for any given mass the gyration radius is smaller at lower metallicity. In addition, we see that for stars with mass $>0.9\rm M_{\odot}$  there is a drop in the gyration radii from 1 Gyr onwards at all metallicities. For smaller mass stars the gyration radius is more stable at least up to 10 Gyr. Most (if not all) previous estimates of orbital stability have been based on the modeled gyration radius at the ZAMS (and solar metallicity) and the current study shows that for any given system the gyration radius is smaller than that at the ZAMS for systems older than 1 Gyr, except for the very low mass primaries where it is similar to the ZAMS value.

Our modelling assumes a fixed mass of the primary. This is unlikely to be the case due to constant mass exchange between the components resulting in constant changing in the gyration radius. This is more likely early in the evolution of contact binaries when mass exchange is thought to be more significant. If one considers the transfer of mass from the secondary to the primary, the increasing mass of the primary will continuously reduce gyration radius and thus the instability mass ratio, (for $M_1 < 1.4 M_\odot$), however, the same transfer of mass to the primary will reduce the mass ratio. Other influences on orbital instability include potential non conservative mass transfer and/or mass loss due to the secondary overflowing its outer Roche lobe. \citet{2022AAS...24030801M} and \citet{2022ApJS..262...12K} have modeled contact binary evolution from formation to orbital instability considering mass transfer profiles from the secondary to the primary as well as the influence of component rotation. Their modelling suggests that instability is likely to occur at low mass ratios but only when the primary rapidly increases its moment of inertia which is most likely in the immediate post main sequence stage. They estimate the instability mass ratio in the range of 0.045 to 0.15.

The end result of the current modelling is that if the primary component of the system is of lower metallicity and we consider it as being halfway through its main sequence lifespan then it is likely to have an instability mass ratio considerably lower than what current practice suggests. For many systems where the mass of the primary is $\geq1\rm M_{\odot}$ the instability mass ratio is so low that the mass of the secondary would fall below the threshold for hydrogen fusion, which is likely to result in considerable increase in energy transfer requirements to maintain thermal equilibrium. Applying a liberal inclusion criteria we find that of the eight systems that satisfy the revised criteria for instability all have methodological or astrophysical issues leading to reduced confidence with respect to their instability status. As noted above, there is only one confirmed merger event. The current and previous work have confirmed orbital instability is likely at low mass ratios with stability improving with reduced metallicity and potentially during the main sequence lifespan. Although hundreds of relatively bright contact binaries have been observed with accurate light curves and spectroscopic solutions the number of identified potential merger candidates is small. We suspect two possible explanations. Firstly, as noted above, contact binaries in the solar neighborhood are metal poor and hence more stable and secondly, we suspect an observational bias of imaging stable systems. Imaging multi-band light curves of contact binaries is a time-consuming exercise and as such only smaller size instruments are available, limiting the observation to relatively bright targets. Brighter targets are more likely to have heavier primaries which are inherently more stable. \citet{2021ApJS..254...10L} catalogued the light curve solutions of over 600 systems with over 400 having an estimate of the mass of the primary component. Less than 150 of these systems have an estimated mass of the primary component below one solar mass. Lower mass contact binaries, which have relatively higher instability mass ratios, are relatively faint and more difficult to identify from the survey photometry and difficult to observe with available instruments. Photometric surveys using large instruments such as the in development Extremely Large Telescope offer an avenue to identify low mass contact binaries.

\section*{Acknowledgements}

\noindent This research has made use of the SIMBAD database, operated at CDS, Strasbourg, France.\\

\noindent N. R. Landin acknowledges financial support from Brazilian agencies FAPEMIG, CNPq, and CAPES.\\



\noindent B. Arbutina acknowledges the funding provided by the Ministry of Science, Technological Development and Innovation of the Republic of Serbia through the contract 451-03-66/2024-03/200104.\\


\noindent During work on this paper, G. Djurašević and J. Petrović were financially supported by the Ministry of Science, Technological Development and Innovation of the Republic of Serbia through contract 451-03-47/2023-01/200002\\

\noindent J. Petrovi\'c, G. Djura\v sevi\'c and B. Arbutina also acknowledge the funding
provided by the Science Fund of the Republic of Serbia through project \#7337 ”Modeling Binary Systems That End in Stellar Mergers and Give Rise to Gravitational Waves” (MOBY).\\

\noindent Based in part on data acquired on the Western Sydney University, Penrith Observatory Telescope. We acknowledge the traditional custodians of the land on which the Observatory stands, the Dharug people, and pay our respects to elders past and present.\\

\section*{Data Availability}
The data underlying this article will be shared on reasonable request to the corresponding author.




\bibliographystyle{mnras}
\bibliography{SSW-bibtex} 





\bsp	
\label{lastpage}
\end{document}